\documentstyle[preprint,aps,epsfig]{revtex}
\tightenlines

\newcommand{\nc}{\newcommand}
\nc{\be}{\begin{equation}}
\nc{\ee}{\end{equation}}
\nc{\bea}{\begin{eqnarray}}
\nc{\eea}{\end{eqnarray}}
\nc{\beas}{\begin{eqnarray*}}
\nc{\eeas}{\end{eqnarray*}}
\nc{\noi}{\noindent}
\nc{\sD}{\not \! \! D}
\nc{\s}[1]{\not \! #1}
\nc{\non}{\nonumber}
\nc{\bb}{\bibitem}
\nc{\lf}{\left}
\nc{\ri}{\right}
\nc{\mb}[1]{\makebox[#1]{}}
\nc{\pa}{\partial}
\nc{\sA}{\not \! \! A}
\nc{\newsec}[1]{\section{#1}\mb{0.5cm}}
\nc{\h}{\frac{1}{2}}
\nc{\ra}{\rightarrow}
\nc{\la}{\leftarrow}
\nc{\etwopi}{$e^+e^-\ra\pi^+\pi^-\;$}
\nc{\ethrpi}{$e^+e^-\ra\pi^+\pi^0\pi^-\;$}
\nc{\lapp}{\hbox{$ {     \lower.40ex\hbox{$<$}
                   \atop \raise.20ex\hbox{$\sim$}
                   }     $}  }
\nc{\rapp}{\hbox{$ {     \lower.40ex\hbox{$>$}
                   \atop \raise.20ex\hbox{$\sim$}
                   }     $}  }
\nc{\M}{{\cal M}}
\nc{\rw}{$\rho$-$\omega\;$}
\def\mathunderaccent#1{\let\theaccent#1\mathpalette\putaccentunder}
\def\putaccentunder#1#2{\oalign{$#1#2$\crcr\hidewidth
\vbox to.2ex{\hbox{$#1\theaccent{}$}\vss}\hidewidth}}

\nc{\ti}{\mathunderaccent\tilde}
\def\hhha{\rule[-3.mm]{0.mm}{7.mm}}

\begin{document}
\preprint{\vbox{                        \hfill UK/TP 97-10 \\
                                        \null\hfill hep-ph/9707385 \\
                                        \null\hfill July 1997
}}
\title{Rho-Omega Mixing and the Pion Form Factor \\ in the Time-like Region}
\author{S. Gardner and H.B. O'Connell}
\address{Department of Physics and Astronomy,
University of Kentucky,\\
Lexington, KY 40506-0055 } 
\maketitle

\begin{abstract}
We determine the magnitude, phase, and $s$-dependence of 
$\rho$-$\omega$ ``mixing'' in 
the pion form factor in the time-like region through fits to 
$e^+e^- \ra \pi^+ \pi^-$ data. 
The associated systematic errors in 
these quantities, arising from the functional form used to fit the 
$\rho$ resonance, are small. The systematic
errors in the $\rho$ mass and width, however, are larger than 
previously estimated. 

\end{abstract}

\narrowtext

\newpage
\section{Introduction}

The pion form factor $F_\pi(s)$ 
in the time-like region is extracted from 
cross section measurements of 
$e^+ e^- \rightarrow \pi^+ \pi^-$, namely, 
\begin{equation}
\sigma(s)=\sigma_{\rm em}(s) |F_\pi(s)|^2 \,, 
\end{equation}
where $\sigma_{\rm em}(s)$ is the cross section to produce
a structureless $\pi^+\pi^-$ pair and $s$ is 
the usual Mandelstam variable. 
The isovector $\rho$ resonance dominates the cross section 
for $\sqrt{s}$ ranging from 600 to 900 MeV, though isospin
violation allows the isoscalar $\omega$ resonance to contribute as well. 
We wish to extract the rho-omega ``mixing'' matrix element 
$\tilde\Pi_{\rho\omega}(s)$, which to 
leading order in isospin violation is given by
\be
F_\pi(s) = F_\rho(s) \left( 
1+\frac{1}{3}\left( \frac{\tilde\Pi_{\rho\omega}(s)}{s - m_\omega^2+
im_\omega\Gamma_\omega} \right) \right) \;, 
\label{fpi}
\ee
from data.
Note that $F_\rho(s)$ is the pion form factor in the absence of isospin
violation. $F_\rho(s)$ is subject only to general theoretical constraints;
it is our purpose to determine how its non-uniqueness 
impacts the extraction of $\tilde\Pi_{\rho\omega}$. Moreover, the $\rho$ 
mass and width are themselves sensitive to the choice of 
$F_{\rho}(s)$~\cite{pdg96,pisut68,ben93}, and we wish to determine the
systematic error in these quantities as well.

Maltman {\it et al.} have discussed the separation of 
$\tilde\Pi_{\rho\omega}(s)$ 
into two contributions: one from the
direct coupling of $\omega \ra 2\pi$ and the other from mixing,
$\omega \ra \rho \ra 2 \pi$~\cite{ber94,malt96,hoc97}. 
Such a separation is model dependent, and we shall not pursue it further. 
Rather, we wish to determine
the constraint $e^+ e^- \ra \pi^+ \pi^-$ data places on the sum of these
contributions; 
we term $\tilde \Pi_{\rho\omega}(s)$ the effective mixing matrix element. 
$\tilde \Pi_{\rho\omega}(s)$ is usually assumed to be both real and 
approximately $s$-independent~\cite{debeau79}; we wish also 
to test
these assumptions
in the context of $e^+ e^- \ra \pi^+ \pi^-$ data. 
An explicit $s$-dependence in $\tilde \Pi_{\rho\omega}(s)$ emerges as 
a consequence of the inclusion of 
a non-resonant contribution to $\omega\ra 2\pi$~\cite{malt96,hoc97}; 
thus, the $s$-dependence
of the effective mixing amplitude in the resonance region 
may partially constrain the role of
this contribution. 

Previous determinations of $\tilde\Pi_{\rho\omega}(s)$ have used
either the empirical $\omega\ra 2\pi$ branching ratio~\cite{gasser82,coon87} 
or $e^+e^- \ra \pi^+ \pi^-$ data~\cite{ber94,hoc95,malt96,hoc97}. 
We prefer the latter method for
several reasons. 
Information on the phase and $s$-dependence of 
$\tilde\Pi_{\rho\omega}(s)$ is not accessible from the 
$\omega\ra 2\pi$ branching ratio. Moreover, the 
determined $\rho$-$\omega$ mixing matrix element then explicitly depends 
on the relatively poorly known $\rho$ resonance parameters. 

Our work differs from earlier analyses in that we enumerate a variety
of forms for $F_{\rho}(s)$ which satisfy the known theoretical constraints.
We extract the magnitude, phase, and $s$-dependence of the effective
\rw mixing matrix element 
$\tilde \Pi_{\rho\omega}(s)$ from \etwopi data and study 
the systematic error in the above parameters resulting from
 the choice of $F_{\rho}(s)$. We have also studied the systematic errors in the
$\rho$ parameter extraction and find them to be much larger than previously
reported. 

To extract $\tilde \Pi_{\rho \omega}(s)$ from \etwopi data we 
naturally wish to use the best data set available. The most recent
data for the pion form factor in the time-like region 
is due to Barkov
{\it et al.}~\cite{barkov85}. They report 
$m_\rho=775.9\pm 0.8 \pm 0.8$ MeV and $\Gamma_\rho=150.5\pm 1.6 \pm 2.5$ MeV,
where the errors respectively refer to the error arising from the
statistical and systematic uncertainties
in the experimental data and to the systematic error resulting from the form
chosen for the pion form factor~\cite{barkov85}.
The Barkov {\it et al.} value for the $\rho$ mass contributes some 40\% of
the $\chi^2$ in the Particle Data Group's 
1994 world average for this quantity~\cite{pdg94}. 
In the 1996 compilation, the Particle Data Group rule the 
Barkov {\it et al.} $\rho$ mass ``probably
wrong'' on statistical grounds and exclude the Barkov {\it et al.} values from 
the world averages for the $\rho$ mass and width~\cite{roos97,pdg96}. 
The only determination of the $\rho$ mass 
currently 
included in the Particle Data Group world average 
based, at least in part, on time-like
pion form factor data 
is that of Heyn and Lang~\cite{hl81}. We
apply our analysis to both the data included in 
Barkov {\it et al.}~\cite{barkov85,wd85} and to
the time-like region data included in Heyn and Lang~\cite{hl81,wd78}, to 
ascertain whether any systematic differences exist between the data
sets. 

The $\rho$ resonance is relatively broad, so that the reported
resonance parameters are numerically sensitive to the convention under
which the mass and width are defined. 
The appearance of a resonance is
associated with a complex pole at $s=s_p$ in the elastic scattering amplitude, 
and this complex pole can be used to define the resonance's 
mass and width~\cite{levy59}. The separation of $s_p$ 
into a mass and width is not unique, however, for both the real
and imaginary parts of $s_p$ appear in any physical process~\cite{stuart95}.
Thus, whether one defines 
$s_p \equiv m^2 - i m \Gamma$, so that $m=\sqrt{{\rm Re}\; s_p}$,
or $\sqrt{s_p} \equiv \tilde m - i \tilde \Gamma/2$, so that 
$\tilde m={\rm Re} \sqrt{s_p}$, is a matter of convention. 
We will present results for the
$\rho$ mass and width under both conventions.

\section{The Pion Form Factor and ${\bf\rho}$ - ${\bf\omega}$ ``mixing''}

Only general theoretical constraints guide the construction 
of the pion form factor in the time-like region.
Charge conservation requires the form factor 
to be unity at $s=0$: 
\be
F_\pi(0)=1 \;.
\label{charge}
\ee 
Moreover, it should be an analytic function in the complex $s$
plane, with a branch cut along the real axis 
beginning at the two-pion threshold, $s=4m_\pi^2$. 
Finally, time-reversal invariance and the unitarity of the $S$ matrix 
requires that the phase of the form factor be that of 
$l=1$, $I=1$ $\pi$-$\pi$ scattering~\cite{fed58}. 
This last emerges as $\pi$-$\pi$
scattering in the relevant channel is very nearly elastic from 
threshold through
$s \approx (m_\pi + m_\omega)^2$~\cite{phase,hl81}. In this region of $s$,
then,
the form factor is related to the $l=1$, $I=1$ phase shift, 
$\delta^1_1$, via~\cite{gas66}
\be\non
F_\pi(s)=e^{2i\delta^1_1}F^*_\pi(s) 
\ee
so that
\be
\tan \delta^1_1(s)=\frac{{\rm Im}\:F_\pi(s)}{{\rm Re}\:F_\pi(s)}\;.
\label{phase}
\ee
The above is a special case of 
what is sometimes called
the Fermi-Watson-Aidzu phase theorem~\cite{gas66,watson55}. 

In the resonance region the phase and analyticity
constraints can be realized via 
the Breit-Wigner form 
\be
\lim_{s\ra m_\rho^2} F_\pi(s) = -
\frac{m_\rho^2(1+\varepsilon)}{s-m_\rho^2+im_\rho
\Gamma_\rho}\;, 
\label{limbw}
\ee
where $\varepsilon$ is a real constant. 
The complex pole $s_p$ associated with the appearance
of the resonance is given by $s_p \equiv m_\rho^2 - i m_\rho \Gamma_\rho$. 
We have adopted the $\sqrt{{\rm Re}\; s_\rho}$ convention for
the $\rho$ mass. Alternatively, one could have written
$\sqrt{s_p} \equiv \tilde m_\rho - i \tilde \Gamma_\rho/2$, so that 
$\tilde m={\rm Re} \sqrt{s_p}$, but this is merely a matter of convention. 
Note that the $m$ and $\tilde m = {\rm Re}\sqrt{s_\rho}$ 
prescriptions are related via~\cite{hl81}
\be
\tilde{m}=\left({\frac{m^2+\sqrt{m^4+m^2\Gamma^2}}{2}}\right)^{\!1/2},\,\,\,\,
\tilde{\Gamma} =\frac{m}{\tilde{m}}\Gamma \;.
\label{reltilde}
\ee

The Breit-Wigner form Eq.~(\ref{limbw}) only satisfies the phase
constraint as $s\ra m_\rho^2$, so that 
a more general form for the pion form factor, suitable for all $s$, is needed. 
Generalizations satisfying the enumerated constraints 
have been constructed by various authors~\cite{gs68,hl81,gesh89}; we
will follow the work of Gounaris and Sakurai~\cite{gs68} and of 
Heyn and Lang~\cite{hl81} in what follows. We include $\rho$-$\omega$
mixing as per Eq.~(\ref{fpi}), so that our enumerated constraints
are brought to bear on the form of $F_\rho(s)$ alone, for the violations
of the above constraints due to isospin breaking are small. 

Gounaris and Sakurai consider a $F_\rho(s)$ of the form $f(0)/f(s)$ 
with $f(s)$ such that 
\be
f(s)=(k^3/\sqrt{s})\cot\delta^1_1-i(k^3/\sqrt{s}) \;, 
\ee
where $f(0)$ is real. 
Note that both the normalization and phase constraints are
manifest in such a construction. The $\delta_1^1$ phase
shift is parametrized via a generalized effective-range formula
of Chew-Mandelstam type~\cite{chew60}, 
with two free parameters, $a'$ and $b'$,
\be
(k^3/\sqrt{s})\cot\delta^1_1=a'+b'k^2 + k^2h(s)\;,
\label{gsscatt}
\ee
where $k$ and $h(s)$ are chosen to be 
\bea\non
k&=&(s/4-m_\pi^2)^{1/2},  h(s)=
\frac{2k}{\pi\sqrt{s}}\log\left(\frac{\sqrt{s}+2k}{2m_\pi}\right) 
 \;;\; s\ge 4m_\pi^2\\
k&=&i(m_\pi^2-s/4)^{1/2},h(s)=\frac{2k i}{\pi\sqrt{s}}
{\rm arccot}\left(\frac{s}{4m_\pi^2-s}\right)^{1/2} \;;\; 0\le s<4m_\pi^2\;.
\label{hdef}
\eea
In this manner the required analytic structure is imposed as well. 
The parameters $a'$ and $b'$ can be replaced by functions of $m_\rho$ and
$\Gamma_\rho$ by noting the resonance conditions
$\cot\delta^1_1(m_\rho^2)=0$ and 
${\delta^1_1}'(m_\rho^2)=1/(m_\rho\Gamma_\rho)$, so that the 
resulting form factor is of the form~\cite{gs68}
\be
F_\rho^{\rm GS}(s)=\frac{-(m_\rho^2+dm_\rho\Gamma_\rho)}{s-
m_\rho^2-\Gamma_\rho (m_\rho^2
/k_\rho^3)[k^2(h-h_\rho)-(s-m_\rho^2)k_\rho^2h'_\rho]+im_\rho\Gamma_\rho(s)
}\;, 
\label{realgsff}
\ee
where
\be
k_\rho=k(m^2_\rho)\;,\quad h_\rho=h(m_\rho^2)\;,\quad 
\Gamma_\rho(s)=\Gamma_\rho\left(\frac{k}{k_\rho}\right)^3
\frac{m_\rho}{\sqrt{s}}\;,
\label{gsff1}
\ee
and $d$ is fixed in terms of $m_\rho$ and $m_\pi$, 
\be
d=\frac{3m_\pi^2}{\pi k_\rho^2}\log\left(
\frac{m_\rho+2k_\rho}{2m_\pi}\right)
+\frac{m_\rho}{2\pi k_\rho}-\frac{m_\rho m_\pi^2}{\pi k_\rho^3}\;.
\label{gsff2}
\ee
Note that the Breit-Wigner form, Eq.~(\ref{limbw}), is recovered 
as $s\ra m_\rho^2$. It turns out that the two-parameter fit of 
Eq.~(\ref{realgsff}) does not 
suffice to fit the pion form factor data in the $\rho$ resonance
region, and a modification of 
$F_\rho^{\rm GS}(s)$ to include a 
multiplicative real function of $s$, constrained by only the
normalization condition, is phenomenologically necessary. 

  It is critical to note that the adoption of an $s$-dependent 
width, $\Gamma(s)$, 
predicated by the phase constraint of Eq.~(\ref{phase}), implies that
$s=m_\rho^2 - i m_\rho\Gamma_\rho$ no longer determines the position of
the complex pole $s_p$~\cite{lang79}. 
Rather, 
if one were to determine the mass and width from 
$s_p\equiv {\overline m}_\rho^2 - 
i {\overline m}_\rho {\overline \Gamma_\rho}$, where $f(s_\rho)=0$, 
then for 
$s={\overline m_\rho}^2$ the phase shift would {\it not} be $\pi/2$, and 
the resultant form
factor near $s={\overline m}_\rho^2$ would {\it not} be of Breit-Wigner form. 
We prefer, then, to determine the resonance parameters 
by comparing
$f(s)$ to $s-m_\rho^2 - i m_\rho \Gamma_\rho$ at the $s$ in the physical
region where the real part of $f(s)$ vanishes. 
The resonance
condition $\cot\delta =\pi/2$ is thus satisfied as $s\rightarrow m_\rho^2$. 
Our procedure is consistent with that of Ref.~\cite{gs68}. 

  Heyn and Lang write $F_\rho(s)$ as 
\be
F_\rho(s) = \Omega(s) F_{\rm red}(s)\;, 
\ee
where for $0 \le s \lapp (m_\omega + m_\pi)^2$, $F_{\rm red}(s)$ is
purely real and can be 
approximated by a cubic polynomial~\cite{hl81},
\be
F_{\rm red}(s)=1+\beta_1 s+\beta_2s^2+\beta_3s^3\;.
\label{fred}
\ee
All phase information is contained in the Omn\`es 
function $\Omega(s)$~\cite{musk53},
which is chosen to be
\be
\Omega(s)=\frac{c+ m_\pi^2 g(0)}{\tilde s_p}\frac{\tilde s_p-s}
{as^2+bs+c-(s-4m_\pi^2)g(s)/4}\;.
\label{omnes}
\ee
Modulo the $(\tilde s_p - s)/\tilde s_p$ factor, $\Omega(s)$ also has
a $f(0)/f(s)$ structure, where now 
\be
f(s)=as^2+bs+c-(s-4m_\pi^2)g(s)/4
\label{hlf}
\ee
and $g(s)$ is determined by the one-pion-loop diagrams in the
$\rho$ self-energy~\cite{lee60,herr93,klingl96},
\bea\non
g(s)&=&
-\frac{1}{\pi}u\log\frac{1+u}{1-u}+iu\;,u=\sqrt{1-4m_\pi^2/s}\;,
s\geq 4m_\pi^2\\
&=&-\frac{2}{\pi}u\arctan\frac{1}{u}\;,u=\sqrt{4m_\pi^2/s-1}\;,0\leq
s\leq4m_\pi^2\label{gdef}\\
&=&-\frac{1}{\pi}u\log\frac{u+1}{u-1}\;,u=\sqrt{1-4m_\pi^2/s}\;,s<0
\non \\
g(0)&=&-2/\pi\;.\non
\eea
The $\rho$ resonance is 
associated with a zero of $f(s)$ in the complex plane for $s>4m_\pi^2$,
yet $f(s)$ also vanishes on
the negative real axis when $s=\tilde s_p$, by definition.  
This zero simulates the left-hand cut in the $N/D$ 
construction of the amplitude~\cite{lang75} and 
physically corresponds to a bound state~\cite{gas66}. The
$(\tilde s_p -s)/\tilde s_p$ factor in 
$\Omega(s)$ removes this singularity while 
preserving the normalization constraint. 
The singularity structure
of $\Omega(s)$ occurs in $F_{\rm GS}(s)$ as well; 
the zero there, though, is at such large negative $s$ that 
the inclusion of a $(\tilde s_p -s)/\tilde s_p$ factor 
would be of no phenomenological impact~\cite{gs68}. 
Two parameters are needed to describe the left-hand cut. 
The third free parameter in Eq.~(\ref{hlf}) functions as 
a CDD parameter; the $N/D$ solution of the partial-wave amplitude
dispersion relation does not uniquely follow from the 
information input on the left-hand cut~\cite{cdd56,gas66}. 

Once $a$, $b$, and $c$ are determined from a 
fit to data, the $\rho$ resonance parameters can also be determined.
As discussed earlier in the case of the Gounaris and Sakurai 
form factor~\cite{gs68}, Eq.~(\ref{realgsff}), we 
determine the resonance parameters by comparing
$f(s)$ to $s-m_\rho^2 - i m_\rho \Gamma_\rho$ at the $s$ in the physical
region where the real part of $f(s)$ vanishes. 
This has the effect of requiring that $f(s)$ be of Breit-Wigner 
form as $s\rightarrow m_\rho^2$. Consequently, 
$m_\rho$
is determined by ${\rm Re}\{f(m_\rho^2)\}=0$ and $\Gamma_\rho$ is
determined by ${\rm Im}\{f(m_\rho^2)\}= - m_\rho \Gamma_\rho$~\cite{hl81}. 

The structure chosen for $f(s)$ in Eq.~(\ref{hlf}) is formally 
consistent with the phase constraint resulting from unitarity
and time-reversal invariance, yet the constraint may not be 
numerically well-satisfied, as the parameters are fit to 
the time-like pion form factor data. 
We can, however, require that the
parameters in $f(s)$ reproduce the empirical $l=1$, $I=1$ $\pi$-$\pi$ 
scattering length $a^1_1$ to gauge whether the results we extract
for the $\rho$ parameters and $\tilde\Pi_{\rho\omega}$ are
sensitive to this additional constraint. We define $a^1_1$ 
as
\be
\frac{1}{a^1_1}=\lim_{s\ra 4m_\pi^2}k^3\cot\delta^1_1\;,
\label{scatt-a}
\ee
where empirically $a^1_1=(0.038\pm0.002)\; m_\pi^{-3}$~\cite{nagels79}.
Note that $1/a^1_1$ is equal to $2m_\pi a'$ in the Gounaris and Sakurai
model, Eq.~(\ref{gsscatt}), and is equal to 
$(16m_\pi^4a+4m^2_\pi b+c)m_\pi$ in the Heyn-Lang model, Eq.~(\ref{omnes}).

The pion form factor we fit to data is 
\be
F_\pi(s)=\Omega(s)F_{\rm red}(s)\left[
1+\frac{1}{3}\frac{\tilde\Pi_{\rho\omega}(s)}{s - m_\omega^2+
im_\omega\Gamma_\omega}\right]\;. 
\label{finfit}
\ee
We work to leading order in isospin violation, and we adopt the
SU(3) value of $1/3$ for the 
ratio of the 
electromagnetic coupling of the $\omega$ to that of the $\rho$. Note that 
$F_{\rm red}(s)$ is given by Eq.~(\ref{fred}) and 
$\Omega(s)$ by 
Eq.~(\ref{omnes}), though we will also replace $\Omega(s)$ by
$F_{\rm GS}(s)$, Eq.~(\ref{realgsff}), and by 
an Omn\`es function modified to resemble $F_{\rm GS}(s)$,
\be
\Omega_{\rm GS}(s)=\frac{c+ m_\pi^2 g(0)}
{bs+c-(s-4m_\pi^2)g(s)/4}\;.
\label{gsff}
\ee
We thus test the sensitivity of the fit to the specific manner
in which the phase constraint is realized. 
Rather than Eq.~(\ref{finfit}), 
Barkov {\it et al.}~\cite{barkov85} adopt a 
pion form factor such that 
\be
F_\pi(s)=F^{\rm GS}_\rho(s)\frac{(1+\alpha_\omega F^{\rm GS}_\omega(s)+
\alpha_{\rho'}F^{\rm GS}_{\rho'}(s)+\alpha_{\rho''}F^{\rm GS}_{\rho''}(s))}{
1+\alpha_\omega+\alpha_{\rho'}+\alpha_{\rho''}}\;,
\label{barkovff}
\ee
where $\alpha_\omega$, $\alpha_{\rho'}$, and $\alpha_{\rho''}$ are 
real constants, to be determined from a fit to data, and 
$F^{\rm GS}_{\rm V}(s)$, with ${\rm V}$ a vector meson, 
is determined by Eqs.~(\ref{realgsff},\ref{gsff1},\ref{gsff2}) 
with $m_\rho,\;\Gamma_\rho$
replaced by $m_{\rm V},\;\Gamma_{\rm V}$. 
$F^{\rm GS}_{\rm V}(s)$ is complex in the region of $s$ we consider, 
so that Eq.~(\ref{barkovff}) would seem to violate the phase constraint
of Eq.~(\ref{phase}). We fit Eq.~(\ref{barkovff}) as well to the time-like
pion form factor data, and we will examine the above issue through
direct comparison with the $l=1$, $I=1$ phase shifts. 

If the inelastic contributions to $\pi$-$\pi$ scattering in the
$s$ region of interest are generated 
by isospin violation only, the solution of the 
Muskhelishvili-Omn\`es integral equation~\cite{musk53} with 
inelastic unitarity suggests that 
the effective $\rho$-$\omega$ mixing matrix element, 
$\tilde \Pi_{\rho\omega}(s)$, 
is real~\cite{debeau79}. It is
also thought to be weakly $s$-dependent~\cite{debeau79}. Thus, unless
otherwise stated, we make the replacement 
$\tilde \Pi_{\rho\omega}(s)=\tilde \Pi_{\rho\omega}(m_\omega^2)$. 
We can, however, also test
these assumptions by replacing $\tilde \Pi_{\rho\omega}(s)$ 
with  
\be
\tilde\Pi_{\rho\omega}(s)\equiv
\tilde\Pi_{\rho\omega}^{\rm R}(m_\omega^2)+i{\rm Im}\:
\tilde\Pi_{\rho\omega}^{\rm I}(m_\omega^2)
\label{piimag}
\ee
or with 
\be
\tilde\Pi_{\rho\omega}(s)=\tilde\Pi_{\rho\omega}(m_\omega^2)+(s-m_\omega^2)
\tilde\Pi'_{\rho\omega}(m_\omega^2) \;,
\label{piener}
\ee
where it is natural to expand the $s$-dependence of the effective
mixing matrix element about $s=m_\omega^2$. In this manner we 
can test whether the value of 
${\rm Re}\{\tilde \Pi_{\rho\omega}(m_\omega^2)\}$ 
is sensitive to the
matrix element's possible $s$ dependence or phase.

\section{Results }\label{Results}

\begin{figure}
\centerline{\epsfig{file=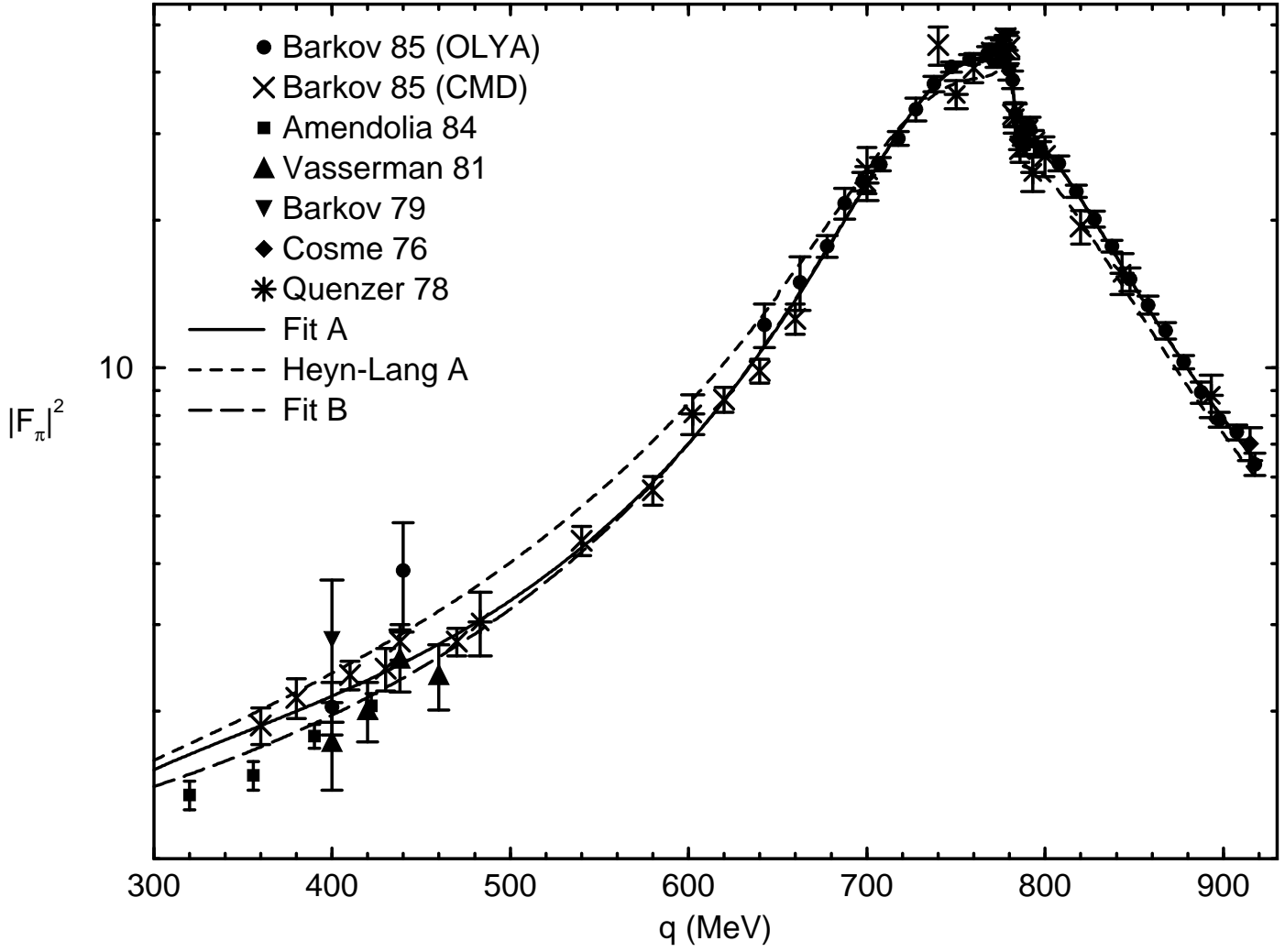,angle=-90,width=5.0in}}
\vspace{100pt}
 \caption{
The absolute square of the time-like pion form factor, 
$|F_\pi(s)|^2$, plotted versus the invariant mass $q$ 
 of the $\pi^+\pi^-$ pair. Fits A (solid line) and B (long-dashed line), 
 as per  Eq.~(\protect{\ref{finfit}}) and Table \protect{\ref{fitdef}},
are shown with
the data compiled by Barkov {\it et al.} The A fit, as per 
Table \protect{\ref{fitdef}}, of Heyn and Lang (dashed line) 
is shown for reference. 
}
 \label{figone}
\end{figure}

We optimize the fit parameters 
using MINUIT~\cite{minuit}. The parameters can be
highly correlated, so that we compute the correlation coefficients
and final errors with a double precision FORTRAN code
using standard techniques~\cite{errors}. 
We fit data from threshold, 
$\sqrt{s}=2m_\pi \sim 280$ MeV, through
$\sqrt{s}\approx m_\omega + m_\pi = 923$ MeV, as this is the region 
in which the 
empirical $l=1$, $\pi=1$ phase shift is essentially elastic~\cite{phase}.
Our data set consists of the 82 points in this energy range included in the 
analysis of Barkov {\it et al.}~\cite{barkov85,wd85}. 
The following fits are 
based on Eq.~(\ref{finfit}) --- the forms chosen 
for $\Omega(s)$, $F_{\rm red}(s)$, and 
$\tilde \Pi_{\rho\omega}(s)$ in each fit are
indicated in Table \ref{fitdef}. 
For fits B and C above, a fit of the type indicated by 
${\rm A}'$ and ${\rm A}^{\rm I}$ has been performed as well, so that
${\rm B}'$ and ${\rm B}^{\rm I}$, {\it e.g.}, 
indicate B-type fits in which $\tilde \Pi_{\rho\omega}(s)$ has been replaced
by Eq.~(\ref{piener}) and Eq.~(\ref{piimag}), respectively. 
Note that in fit B the parameter $c$ is fixed so that the model
reproduces the empirical scattering length, 
$a^1_1=(0.038\pm0.002) \; m_\pi^{-3}$~\cite{nagels79}, as per 
Eq.~(\ref{scatt-a}). 
For definiteness, note that we use $m_\pi=139.57$ MeV, 
$m_\omega=781.94$ MeV, and $\Gamma_\omega=8.43$ MeV~\cite{pdg96} 
in all fits.
The results are given in 
Tables \ref{table1} and \ref{table2}.
 The given parameter errors
arise from the $\chi^2$ optimization of the fit to data, whose errors
include both statistical and experimental systematic 
uncertainties. Specifically, the parameter errors are given by the square root
of the diagonal elements of the inverse of the curvature matrix~\cite{errors}.
The parameters are correlated, so that the errors in the ``Output'' of 
Tables \ref{table1} and \ref{table2} are
generated using the full error matrix. 
All the parametrized forms are able to fit the data 
exceedingly well --- $\chi^2/{\rm dof} \approx 1$ in all cases. 
Fits A and B, along with the Heyn-Lang A-type fit~\cite{hl81} to the 1978 
world data~\cite{wd78}, are plotted as a function of the pion-pair 
invariant mass $q$, $q\equiv \sqrt{s}$, 
with the data~\cite{wd85} 
included in the compilation of Ref.~\cite{barkov85} in
Fig.~\ref{figone}. The shapes of A and B are very similar in the 
resonance region but are visibly different 
at small $q$ as 
fit B is constrained to reproduce the empirical $a_1^1$ scattering length.
The scattering length extracted from fit A is nearly a factor of 2 
larger than the empirical value; later we will explicitly compare the
phase of the fits we generate with the measured 
$l=1$, $I=1$ $\pi$-$\pi$ phase shift
in the $s$ range of interest. 

The effective $\rho$-$\omega$ mixing matrix element 
$\tilde \Pi_{\rho\omega}(m_\omega^2)$ is remarkably insensitive
to the $\rho$ parametrization chosen; 
the value we find is 
$-3500 \pm 300$ ${\rm MeV}^2$. 
This insensitivity 
is significant, for the $\rho$ mass varies by some 10 MeV 
over the same set of the parametrizations. It is likely the consequence
of the narrow $\omega$ width; 
in particular $\Gamma_\rho/\Gamma_\omega \sim 20$~\cite{pdg96}. 
The $s$-dependence and phase of 
$\tilde \Pi_{\rho\omega}(s)$, as per Eq.~(\ref{piener}) and 
Eq.~(\ref{piimag}), are also relatively insensitive to the 
$\rho$ parametrization; this is shown in 
Tables \ref{table1} and \ref{table2}. We conclude that
$\tilde\Pi_{\rho\omega}'$ and $\tilde\Pi_{\rho\omega}^{\rm I}(m_\omega^2)$ 
are $0.03 \pm 0.04$ and $-300 \pm 300\; {\rm MeV}^2$, respectively. 
The errors are such that 
both $\tilde\Pi_{\rho\omega}'$ and 
$\tilde\Pi_{\rho\omega}^{\rm I}(m_\omega^2)$ are consistent with 
zero. This is consistent with the conclusions of 
Costa de Beauregard {\it et al.}~\cite{debeau79}, and thus their implicit
assumptions would seem to be justified in the resonance region. 
Note that the value of ${\rm Re}\{\tilde\Pi_{\rho\omega}(m_\omega^2)\}$
continues to be $-3500 \pm 300\; {\rm MeV}^2$ in the presence of
 $s$-dependent or imaginary contributions to 
$\tilde\Pi_{\rho\omega}(s)$; this is plausible as these
corrections are themselves consistent with zero. Moreover, the
inferred $\rho$ parameters are insensitive to the inclusion of
these effects as well. 

We also consider fits based on the form used by 
Barkov {\it et al.}, Eq.~(\ref{barkovff}).
Fit E uses 
Particle Data Group values for the higher $\rho$ resonances~\cite{pdg96}; 
fit F uses the values adopted by Barkov {\it et al.}~\cite{barkov85}.
These parameters, along with the 
results of fits D, E, and F, are given in Table \ref{tabledef}. 
The values of $\tilde \Pi_{\rho\omega}(m_\omega^2)$ for fits E and F
are determined from the fit to Eq.~(\ref{barkovff}) via the
relation
\be
\tilde\Pi_{\rho\omega}(m_\omega^2) = 
\frac{-3(m_\omega^2 + d_\omega m_\omega \Gamma_\omega)\alpha_\omega}
{1 + \alpha_\omega + \alpha_{\rho'} + \alpha_{\rho''}}  \;,
\label{alpha2pi}
\ee
where $d_\omega$ follows from Eq.~(\ref{gsff2}) with
$m_\rho, \Gamma_\rho \ra m_\omega, \Gamma_\omega$. 
The final error in the 
$\tilde \Pi_{\rho\omega}(m_\omega^2)$ of fits E and F is 
determined by the full error matrix from the fit with 
Eq.~(\ref{barkovff}). The form of Eq.~(\ref{barkovff}) is quite
different from that of Eq.~(\ref{finfit}), yet the values of 
$\tilde \Pi_{\rho\omega}(m_\omega^2)$ are comparable, {\it cf.} 
$-3600 \pm 300\; {\rm MeV}^2$ with $-3500 \pm 300\; {\rm MeV}^2$ from
fits A$-$D. We favor the latter value as the $\rho'$ and $\rho''$
contributions in Eq.~(\ref{barkovff}) serve as 
sources of phase beyond that generated by $F_\rho(s)$. 
We will examine whether such a construction is explicitly at odds with
$\pi$-$\pi$ scattering data. Note, however, that the scattering
lengths extracted from fits D$-$F are reasonably close to
the empirical value, though as $s$ approaches threshold the 
impact of the imaginary part of the $\rho'$ and $\rho''$ 
contributions is minimized. 

The strength of $\rho$-$\omega$ mixing is
commonly extracted from the 
$\omega \ra 2\pi$ branching ratio~\cite{gasser82,coon87}, so that 
it is
useful to reexamine this analysis and compare it with the current
results. 
The physical $|\omega\rangle$ and $|\rho^0\rangle$
are related to the isospin-pure states $|\omega_I\rangle$ and 
$|\rho_I^0\rangle$ via the transformation~\cite{malt96}
\be
\left(
\begin{array}{c}
| \rho_0 \rangle \\
| \omega \rangle \\
\end{array}
\right) 
= 
\left(
\begin{array}{cc}
 1  & -\varepsilon_1 \\
\varepsilon_2 & 1  \\
\end{array} 
\right) 
\left(
\begin{array}{c}
| \rho_I^0 \rangle \\ 
| \omega_I \rangle \\
\end{array}
\right)  
\ee
to leading order in isospin violation. 
Assuming that the vector mesons couple to conserved currents, 
$\varepsilon_1$ and $\varepsilon_2$ 
are determined by requiring that the
physical mixed propagator $D_{\rho\omega}^{\mu\nu}(s)$ 
has no poles, so that 
$\varepsilon_1=\Pi_{\rho\omega}(m_\omega^2 - i m_\omega \Gamma_\omega)/
(m_\omega^2 - m_\rho^2 + i(m_\rho \Gamma_\rho - m_\omega \Gamma_\omega))$
and 
$\varepsilon_2=
\Pi_{\rho\omega}(m_\rho^2 - i m_\rho \Gamma_\rho)/
(m_\omega^2 - m_\rho^2 + i(m_\rho \Gamma_\rho - m_\omega \Gamma_\omega))$.
Then 
\bea
\langle \pi^+ \pi^- | \omega \rangle &=& 
\langle \pi^+ \pi^- | \omega_I \rangle + 
\frac{\Pi_{\rho\omega}(m_\rho^2 - i m_\rho\Gamma_\rho)}
{m_\omega^2 - m_\rho^2 + i(m_\rho \Gamma_\rho - m_\omega \Gamma_\omega)}
\langle \pi^+ \pi^- | \rho_I^0 \rangle \\
&\equiv&
\frac{{\overline{\Pi}}_{\rho\omega}(m_\omega^2)}
{m_\omega^2 - m_\rho^2 + i(m_\rho \Gamma_\rho - m_\omega \Gamma_\omega)}
\langle \pi^+ \pi^- | \rho_I^0 \rangle \;, 
\eea
so that 
\be
\frac{\Gamma(\omega\ra\pi^+\pi^-)}{\Gamma(\rho\ra\pi^+\pi^-)}
= 
\frac{p_\omega^3 m_\rho^2}{p_\rho^3 m_\omega^2}
\left(
\frac{|{\overline{\Pi}}_{\rho\omega} 
(m_\omega^2)|^2}{(m_\rho^2 - m_\omega^2)^2 
+ (m_\rho\Gamma_\rho - m_\omega\Gamma_\omega)^2} \right)
\;, 
\label{pibranch}
\ee
where the overall $p_\omega^3 m_\rho^2/(p_\rho^3 m_\omega^2)$
factor arises from treating $\langle \pi^+\pi^-|\rho_I^0\rangle$, {\it e.g.}, 
as the $g_{\rho\pi\pi}$ coupling as per Ref.~\cite{lee60}.
We assume that $\Gamma(\rho\ra\pi^+\pi^-)=\Gamma_\rho$ and that 
$\langle \pi^+\pi^- | \rho \rangle = \langle \pi^+\pi^- | \rho_I \rangle$,
as corrections to the latter are of nonleading order in isospin violation. 
Using the Particle Data Group values for the 
$\omega\ra 2\pi$ branching ratio, $2.21\pm .30\%$, and for the 
$\rho$ and $\omega$ resonance parameters~\cite{pdg96} 
yields a central value of 
$|{\overline{\Pi}}_{\rho\omega}|=3860 \; {\rm MeV}^2$, 
whereas using the Barkov {\it et al.}
$\omega\ra 2\pi$ branching ratio, $2.3\pm 0.4 \pm 0.2 \%$, and 
$m_\rho=775.9$ MeV and $\Gamma_\rho=150.5$ MeV as per 
their analysis~\cite{barkov85} 
yields 
$|{\overline{\Pi}}_{\rho\omega}|= 3950 \; {\rm MeV}^2$~\cite{picomm}. 
Note that their $\omega\ra 2\pi$ branching ratio is itself extracted 
from Eq.~(\ref{barkovff}), though this and the PDG value use the
estimated 
leptonic widths $\Gamma(\rho\ra e^+e^-)$ and $\Gamma(\omega\ra e^+e^-)$
to infer the ratio of the $\omega$ to $\rho$ electromagnetic couplings,
rather than adopt 
the $SU(3)$ value of $1/3$ as we do~\cite{klingl96}. 
This explains why 
the above results are of slightly larger magnitude than the result for 
$\tilde\Pi_{\rho\omega}$, $-3500 \pm 300\; {\rm MeV}^2$, 
which emerges from the direct analysis of
$e^+e^-\ra \pi^+\pi^-$ data in the time-like region. 
We prefer the latter analysis, however, 
for $\tilde\Pi_{\rho\omega}$ is accessed directly and is
not subject to uncertainties in the $\rho$ parametrization and its
associated parameters.

As the value of the $\rho$ mass extracted by the analysis of 
Barkov {\it et al.} has been called into question on
statistical grounds~\cite{roos97,pdg96}, it is prudent to examine
the sensitivity of our results to the chosen data set. Previously
we have used the 82 data points with $\sqrt{s} \lapp 923 $ MeV 
compiled by Barkov {\it et al.}
and used in their analysis~\cite{wd85}. 
We have 
repeated fits  A$-$C on the 40 points of the 1978 world data~\cite{wd78} 
used by HL~\cite{hl81} and on the 61 data points 
from the OLYA and CMD detectors 
reported by Barkov {\it et al.}~\cite{barkov85} in the $s$ region
of interest. 
Our results are shown in Table \ref{tableall}. Note that the
fit B errors refer to the errors generated by uncertainties in the data
and by the error in the scattering length, respectively. 
We see that the parameters obtained for a given fit are consistent
within errors for the three data sets. Note, moreover, that the 
striking insensitivity 
of $\tilde \Pi_{\rho\omega}(m_\omega^2)$ to the $\rho$ 
parametrization is manifest in all the data sets considered. 

The fits we have considered support a wide range of 
$\rho$ masses and widths. $m_\rho$, for example, ranges from 
763.1 to 774.2 MeV, whereas $\Gamma_\rho$ ranges from 
144.0 to 157.0 MeV. The median values 
are 768.7 and 150.7 MeV respectively, remarkably 
close to the values reported by the Particle Data 
Group~\cite{pdg96}. It is worth noting that the 
two-parameter Gounaris-Sakurai form, realized as either 
fit C, D, E, or F, recalling
Eqs.~(\ref{finfit},\ref{barkovff}) 
and Tables  \ref{fitdef} and \ref{tabledef}),
consistently returns a value of 774 MeV, though fits D, E, and
F all possess $\rho$ widths some 10 MeV smaller than that of fit C. 
Fits C and D only really differ in the manner they parametrize
the $s$-dependence of the phase of the pion form factor, {\it cf.} 
$h(s)$, Eqs.~(\ref{hdef},\ref{gsscatt}), of fit C with 
$g(s)$, Eqs.~(\ref{gdef},\ref{hlf}), of fit D.

\begin{figure}
\centerline{\epsfig{file=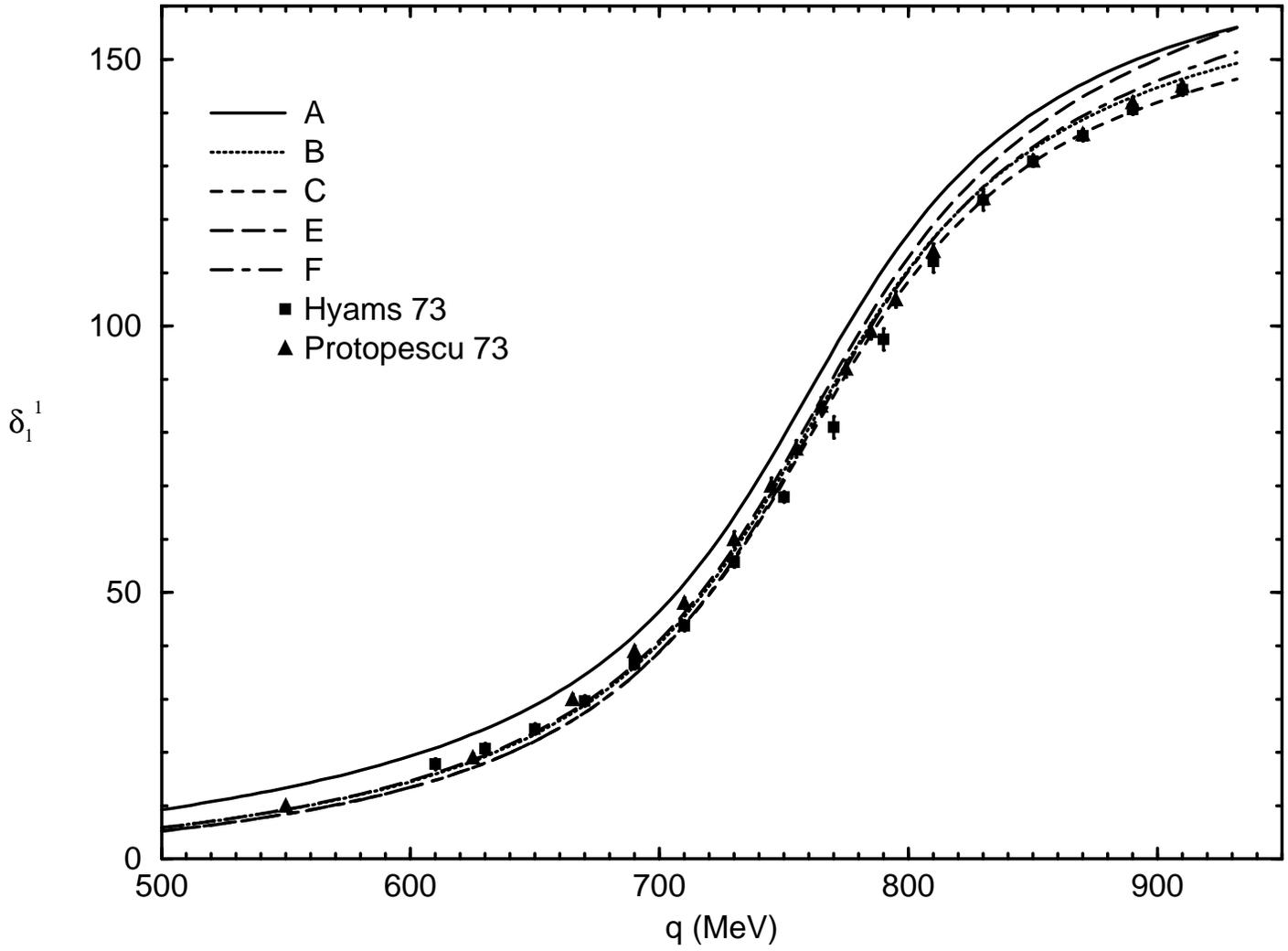,angle=-90,width=5.0in}}
\vspace{100pt}
 \caption{ 
The $l=1$, $I=1$ $\pi$-$\pi$ phase shift $\delta_1^1$ 
extracted from 
the phase of the time-like pion form
factor as determined by fits A (solid line),
B (dotted line), C (dashed line), E (long-dashed line), and 
F (dot-dashed line), given by 
Eqs.~(\protect{\ref{finfit}},
\protect{\ref{barkovff}})
and Tables \protect{\ref{fitdef}}, \protect{\ref{table1}}, 
and \protect{\ref{tabledef}},
plotted versus the pion-pair invariant mass $q$, along with the data from 
Ref.~[26].
Note that the $\rho$-$\omega$ mixing contribution 
to the time-like pion form factor phase has been omitted, 
to facilitate comparison with the empirical phase shifts. 
}
 \label{figtwo}
\end{figure}

The phase
constraint can still be used to discriminate 
between the fits, for the phase of the form factor 
must be numerically that of $l=1$, $I=1$ scattering, if the latter is
elastic. 
We explore this issue in Fig.~\ref{figtwo},
in which the phase of fits A, C, E, and F are shown
as a function of the invariant mass $q$, $q\equiv \sqrt{s}$, 
and compared with data~\cite{phase}.
In plotting the $l=1$, $I=1$ phase shifts, 
we have adopted the energy-independent analysis of
Hyams {\it et al.}~\cite{phase} and the energy-dependent analysis of 
Protopescu {\it et al.}~\cite{phase}.
The above analyses must assume isospin symmetry to separate the
$\delta_l^I$ phase shifts, so that we omit the $\omega$ contribution
from our plots of the phase of $F_\pi(s)$ determined 
from fits to the time-like pion form factor data. 
We have omitted fit D as its phase is essentially identical 
to fit C, which is shown. 
Fits C and D differ in how they parametrize the 
$s$-dependence of the phase of the pion form factor, 
yet this has 
no impact on the extracted $\pi$-$\pi$ phase shift. 
The $\rho$ widths for fits C and D do differ by some $10$ MeV, but this is
of no consequence for the accompanying phase shifts, 
as $\tan^{-1} ({\rm Im} F_\pi/{\rm Re} F_\pi) \ra \pi/2$ as 
$s\ra m_\rho^2$. 
Fits B$-$F are reasonably consistent with data, though to 
judge this in detail we have computed the $\chi^2$ to the
$\pi$-$\pi$ scattering data for the
fits shown in Fig.~\ref{figtwo}. 
\be
\chi^2/{\rm dof} = 1500/34\;[{\rm A}],\,  140/34\; [{\rm B}],\,
140/34\; [{\rm C}],\, 590/34\; [{\rm E}],\,  230/34\; [{\rm F}] \\
\label{chi2}
\ee
With such $\chi^2$'s it is unlikely that any of the time-like 
pion form factor fits are truly correct models of the $\pi$-$\pi$ 
phase shift data, though the employed data sets are themselves
not always consistent within 1$\sigma$. Note, moreover, that
points in the $\rho$-$\omega$ interference region have been included
in the $\chi^2$ computation, though in this region 
the $\omega$ could influence the
phase extracted from the $\rho$ component of the form factor. 
Fit B is decidedly better than fit A; note that B 
differs from A only in that it incorporates the empirical scattering length. 
Fits C, E, and F are also similar in structure; they differ in the 
manner they parametrize the $s$-dependence of the form factor phase --- 
fit C uses $h(s)$, Eqs.~(\ref{hdef},\ref{gsscatt}), whereas fits E and F use
$g(s)$, Eqs.~(\ref{gdef},\ref{hlf}) --- and in that 
the $\rho'$ and $\rho''$ contributions of 
fits E and F, Eq.~(\ref{barkovff}) and Table \ref{tabledef}, 
admit sources of phase
beyond the $\rho$-$\omega$ sector. 
Fit C describes 
the $\pi$-$\pi$ phase shift slightly better than fits E and F. 
As fits C and D yield
phase shifts of comparable shape --- the $\chi^2$/dof for fit D is
also 140/34 ---  it is the presence of the 
$\rho'$ and $\rho''$ contributions in fits E and F 
which is likely responsible for their poorer agreement with data. 
In particular, their inclusion increases the phase shift at large
$q$. We have assumed
that $\delta_1^1$ is purely elastic in
the regime shown, though the measured elasticity $\eta_1^1$ does 
differ slightly from unity above 900 MeV~\cite{phase}. 
The structure of fits E and F 
suggests that this assumption be examined. If the phase shift
is not strictly elastic, that is, if $\eta$
is not exactly unity, then the phase of the form factor $\phi$ is
related to the scattering phase shift $\delta$ via~\cite{debeau79}
\be
\tan \phi \equiv \frac{{\rm Im} F}{{\rm Re} F}
=\frac{(1 - \eta \cos 2\delta)}{\eta \sin 2\delta} \;.
\ee
If $(1 - \eta) \ll 1$, then $\phi$ and $\delta$
are related by 
\be
\phi =\delta + \frac{(1-\eta)\cos \delta}{2\eta \sin\delta} 
\label{inel}
\ee
to leading order in $(1-\eta)$. As $\eta\le 1$ and 
$\delta_1^1 \rapp 100^\circ$ for $q > 800$ MeV~\cite{phase}, 
Eq.~(\ref{inel}) implies that $\phi \le \delta$ as well. Thus 
the structure of fits E and F would seem inconsistent with the manner in
which they fit the phase shifts, for the phase of fits E and F exceeds
that of C for $s\rapp 900 $ MeV, at odds with the constraint of 
Eq.~(\ref{inel}). We conclude that the 
comparison with the measured phase shifts indicates that fits B,C, and D
would seem to be preferred. The average $\rho$ parameters which
emerge from these selected fits are $m_\rho=773 \; {\rm MeV}$ and 
$\Gamma_\rho=153\; {\rm MeV}$. 

There is one last form factor constraint we can consider, for 
the pion form factor is known as $s \ra 0$ from chiral
perturbation theory. Through two-loop order~\cite{gasser91}, 
\be
F_{\pi}(s)= 1 + \frac{1}{6} \langle r^2 \rangle^\pi_V \, s 
+ c_V^{\pi}\, s^2 +  f_V^U (\frac{s}{m_\pi^2}) + 
O(s^3) \;,
\label{chpt}
\ee
where $\langle r^2 \rangle^\pi_V$ 
is the electromagnetic charge radius of the pion 
squared  and $c_V^{\pi}$ is a low-energy constant. 
Note that 
$f_V^U (\frac{s}{m_\pi^2})$ is the genuine loop contribution.
A three-times-subtracted dispersion
integral relates it to $l=1$, $I=1$ $\pi$-$\pi$ scattering
at the tree and one-loop level and to the pion form factor at the 
one-loop level~\cite{gasser91}; a convenient analytic representation
of $f_V^U (\frac{s}{m_\pi^2})$ is given in Ref.~\cite{colangelo96}. 
$f_V^U (\frac{s}{m_\pi^2})$ possesses contributions at both $O(s)$
and $O(s^2)$, yet the separation of the polynomial and dispersive
pieces given in Eq.~(\ref{chpt}) suggests that its contribution to 
 $\langle r^2 \rangle^\pi_V$ is very small --- indeed, this is
numerically the case~\cite{gasser91}. 
Crossing symmetry
dictates the form of the $O(s)$ term and thus offers 
a consistency check of the space-like and time-like region data. 
$\langle r^2 \rangle^\pi_V$ 
has been measured in the space-like 
region by Amendolia {\it et al.}; adopting a fit in which 
$|F_\pi(t)|^2 = n/(1 - t\cdot \frac{1}{6}\langle r^2\rangle^\pi_V)^2$ and
$F_\pi(0)=n=1.000 \pm 0.009$  
yields 
$\langle r^2 \rangle^\pi_V
=0.431 \pm 0.010 \; {\rm fm}^2$~\cite{amendolia86}. 
The data of Amendolia {\it et al.} has been reanalyzed by
Colangelo {\it et al.} using the form dictated by chiral perturbation
theory, Eq.~(\ref{chpt}), and they find 
\bea
\langle r^2 \rangle^\pi_V &=& 0.431 \pm 0.020 \pm 0.016 \; {\rm fm}^2 \\
c^\pi_V &=& 3.2 \pm 0.5 \pm 0.9 \; {\rm GeV}^{-4} 
\eea
where the errors refer to statistical and theoretical uncertainties, 
respectively. 
The agreement of $\langle r^2 \rangle^\pi_V$ with the value
extracted by Amendolia {\it et al.} also indicates that the 
contribution of $f_V^U(\frac{s}{m_\pi^2})$ at $O(s)$ is small. 
The larger statistical error is a consequence of the two-parameter
fit. 

In the time-like region, $\langle r^2 \rangle^\pi_V$ 
is determined by fits to data above the two-pion threshold, $s \ge 4\pi^2$. 
Ignoring the negligible $\tilde\Pi_{\rho\omega}$ contribution, 
one has for fits A and B, {\it e.g.},  that 
\be
\langle r^2 \rangle^\pi_V = 6
\left( \beta_1 - 
\frac{1}{s_p} - \frac{1}{3} \left(
\frac{3b\pi + 2}{c\pi -2 m_\pi^2}
\right)
\right) \;, 
\ee
whereas the $\langle r^2 \rangle$ contribution from $F_\rho^{\rm GS}(s)$,
Eqs.~(\ref{realgsff},\ref{gsff1},\ref{gsff2}), is 
\be
\langle r^2 \rangle 
= \frac{6}{m_\rho^2 + d m_\rho \Gamma_\rho} 
\left( 
1 + \frac{1}{2\pi} \frac{\Gamma_\rho}{k_\rho}
+ \frac{\Gamma_\rho m_\rho}{2\pi k_\rho^2} 
\left( 1 + 2\frac{m_\pi^2}{m_\rho^2} \right)
\log\left(\frac{m_\rho + 2 k_\rho}{2m_\pi}\right)
- \frac{1}{3\pi} \frac{\Gamma_\rho m_\rho^2}{k_\rho^3}
\right)
\;.
\ee
Using the parameters of Table \ref{table1} and \ref{tabledef}, one finds 
\be
\langle r^2 \rangle^\pi_V
\; ({\rm fm}^2) = 0.30 \pm 0.04\; [{\rm A}],\, 
 0.35 \pm 0.03 \pm 0.02\; [{\rm B}],\,  
0.36 \pm 0.03\; [{\rm C} \& {\rm D}],\,  
0.40 \pm 0.01\; [{\rm E} \& {\rm F}]. 
\label{charr}
\ee
Fit A does poorly on the $a_1^1$ scattering length, so that its 
disagreement with the value of $\langle r^2 \rangle^\pi_V$ from space-like
data is not surprising. 
Fit B, however, is constrained to
reproduce $a_1^1$ and is much closer to the Colangelo {\it et al.} 
value~\cite{colangelo96}. Note that the errors associated with fit
B are to be added in quadrature; the first error arises from the
error in the fits, the second from the error in $a_1^1$. The
agreement of fits C and D is comparable to that of fit B, whereas the
results of fits E and F are within error of the Colangelo 
{\it et al.} value, when its error is taken into account as well. 
Thus, we see no real evidence for disagreement with the space-like data,
and its consideration does not serve to distinguish the fits B$-$D,
preferred by the phase shifts. 

A plurality of conclusions exists in the literature concerning the 
consistency of the space-like and time-like 
data~\cite{hl81,alonso87,dubnicka89,gesh89}. Of those who have
argued that they are inconsistent~\cite{hl81,alonso87}, the difficulty
seems to be with data at larger $t$~\cite{hl81,dubnicka89}, well beyond the
region from which the charge radius is extracted. 

Let us conclude this section by summarizing our results for the
$\rho$ mass and width. Fits A$-$D are all distinct in character and
generate excellent fits to the time-like pion form factor data. 
Presuming that the central values are normally distributed, we
can compute the variance of the central values to infer the 
theoretical systematic error in a particular parameter. Using the
$m$ prescription, so that the form factor is given by Eq.~(\ref{limbw})
as $s \rightarrow m_\rho^2$,
averaging over the results of fits A$-$D yields, 
\be
m_\rho = 770.5 \pm 1.1 \pm 4.4 \; {\rm MeV} \;,
\qquad
\Gamma_\rho= 153.5 \pm 0.9 \pm 4.0 \; {\rm MeV} \;,
\label{rhopar}
\ee
where the errors reflect the experimental statistical and
systematic uncertainties and the theoretical systematic 
uncertainties
arising from the $\rho$ parametrization chosen, respectively.
Note that we have discarded the error arising from the use of the 
empirical scattering
length in Fit B in the above, 
as realistically one would refit the parameters
should the scattering length be varied. 
These values differ from those of
Barkov {\it et  al.},
$m_\rho = 775.9 \pm 0.8 \pm 0.8$ MeV and
$\Gamma_\rho = 150.5 \pm 1.6 \pm 2.5$ MeV~\cite{barkov85}, 
though the most significant 
difference is in our respective estimates of the model errors, 
{\it cf.} $4.4$ MeV with $0.8$ MeV for the $\rho$ mass and 
$4.0$ MeV with $2.5$ MeV for the $\rho$ width. 
The theoretical systematic error we report is sensitive to 
the manner in which we determine the $\rho$ mass and width. 
If, alternatively, we compute the mass and width from the value of the
complex pole $s_p$, namely $s_p \equiv {\overline m}^2 - 
i {\overline m}{\overline \Gamma}$, we find for fits 
A$-$D that ${\overline m}_\rho$ ranges from 756 to 757 MeV and 
${\overline \Gamma}_\rho$ ranges from 141 to 142 MeV. A
similar parametrization insensitivity 
of the mass and width defined from 
the complex pole position has been noted in Ref.~\cite{ber94}. 
Note, however, for this latter choice, that
the phase shift would not be $\pi/2$ at $s={\overline m_\rho}^2$ 
and that
the resultant form
factor near $s={\overline m}_\rho^2$ would not be of Breit-Wigner form. 
Determining the mass and width as we have previously, but 
adopting 
the $\tilde m$ prescription, noting Eq.~(\ref{reltilde}),
fits A$-$D yield
\be
\tilde m_\rho = 774.3 \pm 1.1 \pm 4.4 \; {\rm MeV} \;,
\qquad
\tilde \Gamma_\rho= 152.7 \pm 0.9 \pm 3.9 \; {\rm MeV} \;.
\ee
If we discard fit A for its poor fit to the $\pi$-$\pi$ phase shifts and
repeat the above procedure for fits B$-$D only, we find 
\be
m_\rho = 773.0 \pm 0.7 \pm 1.2 \; {\rm MeV} \;,
\qquad
\Gamma_\rho= 153.3 \pm 1.1 \pm 4.6 \; {\rm MeV} 
\ee
and
\be
\tilde m_\rho = 776.8 \pm 0.7 \pm 1.1 \; {\rm MeV} \;,
\qquad
\tilde \Gamma_\rho= 152.6 \pm 1.1 \pm 4.5 \; {\rm MeV} \;. 
\ee
This step-wise elimination procedure may not be reasonable, however, as
slight adjustments of fit A may yield an acceptable 
simultaneous description of the $l=1$, $I=1$ 
$\pi$-$\pi$ phase shift and time-like pion form factor data 
with comparable values of the $\rho$ 
resonance parameters. 
We favor our determination based on the time-like form factor data alone,
Eq.~(\ref{rhopar}).

\section{Conclusions}

We have determined the magnitude, phase, and $s$-dependence 
of the effective $\rho$-$\omega$ mixing matrix element 
$\tilde \Pi_{\rho\omega}(s)$ from fits to 
$e^+e^- \ra \pi^+ \pi^-$ data in the context of a 
theoretical framework which satisfies the 
analyticity, normalization, and phase constraints. 
We have considered a variety of descriptions which satisfy these 
theoretical constraints and have found the $\rho$-$\omega$ mixing
parameters of interest to be insensitive to the manner in which the
$\rho$ is parametrized.  Empirically 
$\Gamma_\omega \ll \Gamma_\rho$, and this likely drives the above
result. Recalling Eqs.~(\ref{piimag},\ref{piener}),
we find 
\bea\non
\tilde \Pi_{\rho\omega} (m_\omega^2) &=& -3500 \pm 300\; {\rm MeV}^2 \\
\tilde\Pi_{\rho\omega}^{\rm I}(m_\omega^2) &=& -300 \pm 300 \; {\rm MeV}^2 \\
\tilde \Pi_{\rho\omega}'  &=& 0.30 \pm 0.40 \; {\rm MeV}^2 \;, \non
\eea
where $\tilde\Pi_{\rho\omega}^{\rm I}(m_\omega^2)$ denotes the
imaginary part of the effective mixing matrix element at $s=m_\omega^2$
and $\tilde \Pi_{\rho\omega}'$ characterizes the $s$-dependence
of $\tilde\Pi_{\rho\omega}(s)$ about $s=m_\omega^2$. Both the
phase and $s$-dependence of $\tilde \Pi_{\rho\omega}(s)$ are statistically
insignificant. It is not that these effects are 
numerically trivial, but rather that they are poorly constrained by 
current $e^+e^- \ra \pi^+ \pi^-$ data. 

 The value of $\tilde \Pi_{\rho\omega}(m_\omega^2)$ we extract 
differs slightly from that
extracted from  the $\omega\ra 2\pi$ branching ratio, 
Eq.~(\ref{pibranch}), that is,
$|\tilde \Pi_{\rho\omega}(m_\omega^2)| \approx 3900\, - \, 4000 \;
{\rm MeV}^2$. 
$\Gamma(\omega\ra 2\pi)$ is itself extracted from the
time-like pion form factor data, and its connection to 
$\tilde\Pi_{\rho\omega}(m_\omega^2)$ 
explicitly involves the
$\rho$ parameters, which are relatively uncertain. 
It thus seems more appropriate 
to extract $\tilde\Pi_{\rho\omega}(m_\omega^2)$ 
directly from $e^+e^-$ data, as we have done. 
Were we to use Eq.~(\ref{pibranch}) to extract 
${\rm B}(\omega\ra 2\pi)$ from our value of 
$\tilde\Pi_{\rho\omega}(m_\omega^2)$, not only would
it be smaller than that which 
Barkov {\it et al.} reports~\cite{barkov85}, 
but it would also be explicitly sensitive to the $\rho$ parametrization.

We have also systematically explored the $\rho$ parameters associated
with our parametrizations of the pion form factor. We find that the
time-like data supports a range of $\rho$ parameters. Adopting the 
prescription currently favored by the
Particle Data Group~\cite{pdg96}, we find 
\be
m_\rho = 770.5 \pm 1.1 \pm 4.4 \; {\rm MeV} \;,
\qquad
\Gamma_\rho= 153.5 \pm 0.9 \pm 4.0 \; {\rm MeV} \;,
\ee
where the errors refers to the empirical and theoretical 
systematic uncertainties, respectively. 
It is worth noting that the differing prescriptions for the mass
and width 
plague the direct comparison of the results of different groups, for the
most recent world average for the $\rho$ mass~\cite{pdg96} 
mixes results defined under different conventions~\cite{examples}.
The convention favored by the Particle Data Group has itself changed
with time, {\it cf.} the $\tilde m$
convention, noting Eq.~(\ref{reltilde}), of Ref.~\cite{pdg78}
with the $m$ convention, noting Eq.~(\ref{limbw}),
of Ref.~\cite{pdg96}, with no apparent 
adjustment of the reported values. 
We agree with the $\rho$ 
width extracted by Barkov {\it et al.},
$\Gamma_\rho = 150.5 \pm 1.6 \pm 2.5$ MeV, but their value
for the $\rho$ mass, 
$m_\rho = 775.9 \pm 0.8 \pm 0.8$ MeV, 
seems slightly large~\cite{barkov85}. It is the latter value that
the Particle Data Group excludes on 
statistical grounds~\cite{pdg96,roos97}.
Note that our theoretical systematic errors
are much larger than those reported by Barkov {\it et al.}~\cite{barkov85},
{\it cf.} $4.4$ MeV with $0.8$ MeV for the $\rho$ mass. 
We have also studied the $\rho$ parameters which result from
different choices of data sets and find that the parameters extracted
from the data of Barkov {\it et al.} are merely more precise. We conclude,
then, that the data of Barkov {\it et al.}~\cite{wd85} 
are not likely wrong, 
but rather that 
the theoretical systematic errors associated
with the extraction of the $\rho$ parameters are larger than
previously estimated.

\begin{center}
{\bf Acknowledgements}
\end{center}
We thank W.~Korsch for many discussions, much 
helpful advice, and for comments on the manuscript. 
We are grateful to C.B. Lang for promptly
communicating the data included in the fits of Ref.~\cite{hl81}
and to C.J. Horowitz and A.W. Thomas for helpful remarks. 
S.G. thanks U.-G. Mei$\ss$ner for suggesting the use of Ref.~\cite{gasser91}
and for helping us obtain the data of Hyams {\it et al.}~\cite{phase},
and H.O.C. thanks  M.~Benayoun and M.J.~Peardon
for helpful discussions. This work was supported by the U.S.
Department of Energy under Grant DE--FG02--96ER40989.


\begin{table}[htb]
\begin{center}
\caption{Components of $F_\pi(s)$ for each fit. Note that ``$-$'' means 
$\tilde \Pi_{\rho\omega}(s)=\tilde \Pi_{\rho\omega}(m_\omega^2)$. 
}
\begin{tabular}{lccc}
Fit & $\Omega(s)$ & $F_{\rm red}$ &$ \tilde\Pi_{\rho\omega}(s)$ \\
\hline
A & $ {\rm Eq.}~(\ref{omnes})$ & ${\rm Eq.}~(\ref{fred})$ & $-$\\
${\rm A}'$ &  ${\rm Eq.}~(\ref{omnes})$ &$ {\rm Eq.}~(\ref{fred})$ &
{Eq.}~(\ref{piener}) \\
${\rm A}^{\rm I}$ &$ {\rm Eq.}~(\ref{omnes})$ &$ {\rm Eq.}~(\ref{fred})$ &
${\rm Eq.}~(\ref{piimag})$ \\
{B} &  ${\rm Eq.}~(\ref{omnes}) \;( c\;{\rm fixed})$ &
{Eq.}~(\ref{fred}) & $-$ \\
{C} &  {Eq.}~(\ref{gsff}) & {Eq.}~(\ref{fred}) & $-$ \\
{D} &  {Eq.}~(\ref{realgsff}) & {Eq.}~(\ref{fred}) & $-$ \\
\end{tabular}
\label{fitdef}
\end{center}
\end{table}

\begin{table}[htb]
\begin{center}
\caption{
Results from fitting the data of the 
Barkov {\it et al.}
compilation
from threshold through 923 MeV with 
$F_\pi(s)$ 
as per Eq.~(\protect{\ref{finfit}}) and Table \protect{\ref{fitdef}}. 
Note that $m_\pi=0.13957$ GeV.
In fit B, the first set of errors associated with the $\rho$
resonance parameters is determined by the statistical and
experimental systematic errors in the time-like pion form factor data,
whereas the second set of errors arises from the error in the
input empirical scattering length. 
$^*$Not a fitting parameter. $^\dagger$Input. 
}

\begin{tabular}{cccccc}
\hhha {Parameter} & A & B & C & ${\rm A}'$ & ${\rm A}^{\rm I}$ \\
\hline
$10^2\cdot a\;(m_\pi^{-2})$ 
& $-2.80\pm 0.88$ & $-0.672\pm 0.084$ & $-$ & 
$-2.68\pm 0.95$ & $-2.57\pm 1.06$\\
$b$ 
& $0.24\pm0.52$ & $-1.008\pm0.027$ & $-1.403\pm0.023$
& $0.17\pm0.56$ & $0.10\pm0.63$\\
$c\;(m_\pi^{2})$ 
& $11.4\pm8.1$ & $30.45^*$ & $36.48\pm0.66$
&$12.5\pm8.7$ & $13.7\pm9.6$\\
$\beta_1$ (GeV$^{-2}$) 
& $-0.65\pm0.25$ & $-0.27\pm0.12$ & $-0.16\pm0.12$
& $-0.69\pm0.25$ & $-0.57\pm0.27$\\
$\beta_2$ (GeV$^{-4}$) 
& $1.80\pm0.91$ & $0.76\pm0.46$ & $0.65\pm0.48$ 
& $1.85\pm 0.88$  & $1.53\pm 0.93$ \\
$\beta_3$ (GeV$^{-6}$) 
& $-0.97\pm0.68$ & $-0.34\pm0.40$ & $-0.29\pm0.42$ 
&$-1.01\pm0.66$ & $-0.79\pm0.68$\\
$\tilde\Pi_{\rho\omega}$ (MeV$^2$) 
& $-3460\pm290$ & $-3460\pm290$ & $-3460\pm 290$
& $-3500\pm300$ & $-3480\pm280$ \\
$\tilde\Pi'_{\rho\omega}$ 
&$-$&$-$&$-$&$0.027\pm0.040$ &$ -$\\
Im $\tilde\Pi_{\rho\omega}$ (MeV$^2$) 
&$-$&$-$&$-$&$-$ &$-310\pm280 $\\
\hline
$\chi^2$/dof & 68/75 & 68/76 &68/76 & 67/74 & 66/74\\
\hline
Output &&&\\
\hline
$\tilde{s}_p\;(m_\pi^2)$ & $-12.46$ & $-125.7$ & $-$ & $-14.21$ & $-8.214$\\
$a^1_1\;(m_\pi^{-3})$ & $0.084\pm0.043$ & $0.038\pm 0.002^\dagger$ & 
$0.0324\pm0.0024$ & 
$0.079\pm 0.041$ & $0.073\pm0.039$\\
$m_\rho$ (MeV) & $763.1\pm 3.9$ & $771.3\pm1.3\pm16$ & $773.9\pm1.2$
& $763.7\pm4.1$ & $764.5\pm4.5$\\
$\Gamma_\rho$ (MeV) & $153.8\pm1.2$ & $156.2\pm0.4\pm4.7$ & $157.0\pm0.4$
& $154.0\pm 1.2$ & $154.2\pm1.3$\\
$\tilde{m}_\rho$ (MeV) & $766.9\pm 4.0 $ & $775.2\pm1.3\pm16$  & $777.8\pm1.2$
& $767.5\pm4.2$ & $768.3\pm4.5$ \\
 $\tilde{\Gamma}_\rho$ (MeV) & $153.0\pm1.2$ &$155.4\pm0.4\pm4.7$
 & $156.2\pm0.4$
 & $153.2\pm1.2$ & $153.4\pm 1.3$ \\
\end{tabular}

\label{table1}
\end{center}
\end{table}

\begin{table}[htb]
\begin{center}

\caption{
More results from fitting the data of the 
Barkov {\it et al.} compilation
from threshold through 923 MeV with 
$F_\pi(s)$ 
as per Eq.~(\protect{\ref{finfit}}) and Table \protect{\ref{fitdef}}. 
Note that $m_\pi=0.13957$ GeV.
Here the sensitivity of the 
$s$-dependence and phase of 
$\tilde\Pi_{\rho\omega}(s)$ to the
$\rho$ parametrization 
is examined. 
$^*$Not a fitting parameter. $^\dagger$Input. 
}

\begin{tabular}{ccccc}
\hhha {Parameter}     & ${\rm B}'$ & ${\rm B}^{\rm I}$ & 
 ${\rm C}'$ & ${\rm C}^{\rm I}$  \\
\hline
$10^2\cdot a\;(m_\pi^{-2})$  & $-0.690\pm 0.087$ & $-0.712\pm0.090$ 
                      & $-$               & $-$               \\
$b$                   & $-1.002\pm0.028$  & $-0.993\pm0.030$  
                      & $-1.408\pm0.023$  & $-1.413\pm0.024$  \\
$c\;(m_\pi^{2})$        & $30.43\pm0.10^*$    & $30.40\pm0.10^*$   
                      & $36.62\pm0.68$    & $36.80\pm0.71$    \\
$\beta_1$ (GeV$^{-2}$)& $-0.35\pm0.15$   & $-0.25\pm0.12$   
                      & $-0.25\pm0.16$  & $-.14\pm0.13$   \\
$\beta_2$ (GeV$^{-4}$)& $0.93\pm0.49$    & $0.68\pm0.46$    
                      & $0.82\pm0.51$   & $0.56\pm0.48$    \\
$\beta_3$ (GeV$^{-6}$)& $-0.45\pm0.42$   & $-0.28\pm0.40$   
                      & $-0.40\pm0.43$  & $-0.22\pm0.42$   \\
$\tilde\Pi_{\rho\omega}$ (MeV$^2$)&$-3510\pm 300$& $-3480\pm290$     
                      & $-3510\pm300$     & $-3480\pm290$     \\
$\tilde\Pi_{\rho\omega}'$&$0.034\pm0.039$     & $-$               
                      & $0.035\pm0.039$ & $-$               \\
Im $\tilde\Pi_{\rho\omega}$ (MeV$^2$)&$-$         & $-340\pm270$        
                      & $-$               & $-340\pm260$      \\
\hline
$\chi^2$/dof          &  67/75         & 66/75            
                      & 67/75          & 66/75          \\
\hline
Output &&&& \\
\hline
$\tilde s_p\;(m_\pi^2)$        & $-122.72$          & $-119.01$          
                      & $-$               & $-$               \\
$a^1_1\;(m_\pi^{-3})$   & $0.038\pm0.002^\dagger$ & $0.038\pm0.002^\dagger$ 
                      & $0.0323\pm0.0024$   & $0.0321\pm0.0025$   \\
$m_\rho$ (MeV)        & $771.3\pm1.3\pm15$& $771.5\pm1.3\pm15$
                      & $773.9\pm1.2$     & $774.1\pm1.2$     \\
$\Gamma_\rho$ (MeV)   &$156.2\pm0.4\pm4.3$& $156.3\pm0.4\pm4$ 
                      & $157.0\pm0.4$     & $157.0\pm0.4$     \\
$\tilde m_\rho$ (MeV)&$775.2\pm1.3\pm15$    & $775.4\pm1.3\pm15$
                      & $777.8\pm1.2$     & $778.1\pm1.2$     \\
$\tilde \Gamma_\rho$ (MeV)&$155.4\pm0.4\pm4.3$&$155.5\pm0.4\pm4$
                      & $156.2\pm0.4$     & $156.3\pm0.4$     \\
\end{tabular}

\label{table2}
\end{center}
\end{table}

\begin{table}[htb]
\begin{center}

\caption{
Results from fitting the data of the 
Barkov {\it et al.} compilation
from threshold through 923 MeV with 
$F_\pi(s)$ 
as per Eq.~(\protect{\ref{barkovff}}) and 
Eqs.~(\protect{\ref{realgsff}},\protect{\ref{gsff1}},\protect{\ref{gsff2}}). 
However, note that fit D, as per Table
\protect{\ref{fitdef}}, uses Eq.~(\protect{\ref{finfit}}) with 
Eqs.~(\protect{\ref{realgsff}},\protect{\ref{fred}}) in place of 
Eq.~(\protect{\ref{barkovff}}).
For Fit D $\beta_1 = -0.16 \pm 0.12 \; {\rm GeV}^{-2}$, 
$\beta_2 = 0.65 \pm 0.48 \; {\rm GeV}^{-4}$, and 
$\beta_3 = -0.29 \pm 0.42 \; {\rm GeV}^{-6}$.
The values of $\tilde \Pi_{\rho\omega}(m_\omega^2)$ for fits E and F
are determined from the fit to Eq.~(\protect{\ref{barkovff}}) via 
Eq.~(\protect{\ref{alpha2pi}}).
$^*$Not a fitting parameter. $^\dagger$Input. 
}

\begin{tabular}{cccc}

\hhha Parameter & D &  E &  F \\
\hline
 $m_{\rho}$ (MeV)  & $773.9 \pm 1.2$ & $774.2 \pm 1.2$ & 
$773.8 \pm 1.1$ \\ 
 $\Gamma_{\rho}$ (MeV)  & $146.9 \pm 3.4$ & $145.7 \pm 2.2$ &
$144.0 \pm 2.3$ \\
 $\alpha_\omega$  & $-$ & $(1.20 \pm 0.34)\cdot 10^{-3}$  &
$(1.50 \pm 0.28) \cdot 10^{-3}$ \\
 $\alpha_{\rho'}$ & $-$ & $1.01 \pm 0.18$ &
$0.467 \pm 0.096$ \\ 
 $\alpha_{\rho''}$ & $-$ & $-1.40 \pm 0.33$ &
$-0.70 \pm 0.20$ \\ 
$ m_{\rho'}\;({\rm MeV})$ & $-$ & 1465$^\dagger$ &  1290$^\dagger$ \\
$\Gamma_{\rho'}\;({\rm MeV})$ & $-$ & 310$^\dagger$ & 200$^\dagger$ \\
$ m_{\rho''}\;({\rm MeV})$ & $-$ & 1700$^\dagger$ & 1590$^\dagger$ \\
$\Gamma_{\rho''}\;({\rm MeV})$ & $-$ &235$^\dagger$ & 260$^\dagger$\\
\hline
$\chi^2/{\rm dof}$ & $68 / 76$ & $74 / 77$  &
$71 / 77$ \\
\hline
$\tilde \Pi_{\rho \omega}\;({\rm MeV}^2)$
& $-3460 \pm 290$ & $-3610 \pm 310^*$ & $-3580 \pm 310^*$ \\
$\tilde m_{\rho}\;({\rm MeV})$ & $777.3 \pm 1.2$
 & $777.6 \pm 1.2$ &
$777.1 \pm 1.1$ \\
 $\tilde \Gamma_{\rho}\;({\rm MeV})$ & $146.2 \pm 3.3$ &  
$145.1 \pm 2.2$
& $143.4 \pm 2.2$ \\
 $a_1^1\;(m_\pi^{-3})$ & $(3.240 \pm 0.060)\cdot 10^{-2}$
& $(3.208 \pm 0.036)\cdot 10^{-2}$ &
$(3.182 \pm 0.038)\cdot 10^{-2}$  \\ 
\end{tabular}

\label{tabledef}
\end{center}
\end{table}

\begin{table}[htb]
\begin{center}
\caption{
$\rho$ parameters and 
$\tilde \Pi_{\rho\omega}(m_\omega^2)$ resulting from 
fits A$-$C of 
Table \protect{\ref{fitdef}} and 
Eq.~(\protect{\ref{finfit}}) to the time-like pion form 
factor data, \protect{$|F_\pi(q^2)|^2$}. ``40'' denotes
the 40 points of the 1978 world data. ``60'' denotes the
OLYA and CMD data of Barkov {\it et al.}, whereas 
``82'' denotes the data set compiled by Barkov {\it et al.}
}
\begin{tabular}{ccccc}
\hhha {Fit} & $m_\rho$ & $\Gamma_\rho$ & $\tilde\Pi_{\rho\omega}$ &
$\chi^2$/dof \\
\hline
A40 & $768\pm12$  & $155.3\pm3.5$ & $-2970\pm690$ & 41/33 \\
A61 & $759.7\pm4.1$ & $152.8\pm1.2$ & $-3340\pm300$ & 38/54 \\
A82 & $763.1\pm3.9$ & $153.8\pm1.2$ & $-3460\pm290$ & 68/75 \\
\hline
B40 & $769.7 \pm 3.6 \pm 15.5$  & $155.7\pm 1.1 \pm 4.5$ 
& $-2970\pm690$ & 41/34\\
B61 & $771.7 \pm 1.3 \pm 15.6$  & $156.3\pm 0.38 \pm 4.6$ 
& $-3310\pm300$ & 43/55 \\
B82 & $771.3 \pm 1.3 \pm 16.1$  & $156.2\pm 0.37 \pm 4.7$ 
& $-3460\pm290$ & 68/76 \\
\hline
C40 & $772.4\pm3.6$ & $156.5\pm1.0$ & $-2970\pm690$ & 41/34 \\
C61 & $774.0\pm1.2$ & $157.2\pm0.4$ & $-3310\pm300$ & 43/55 \\
C82 & $773.9\pm1.2$ & $157.0\pm0.4$ & $-3460\pm290$ & 68/76 \\
\end{tabular}
\label{tableall}
\end{center}
\end{table}


\end{document}